\documentclass[10pt]{article}
\usepackage{graphicx}
\usepackage{caption}
\usepackage{subcaption}
\begin{document}

\title{\bf  Quantifying the uncertainty in the time-redshift relationship}
\author{Michael S. Turner \\
Kavli Institute for Cosmological Physics\\
University of Chicago, Chicago, IL  60637-1433\\
\\
Department of Physics and Astronomy\\
The University of California, Los Angeles\\
Los Angeles, CA  90095-1547 \\
\\
email: mturner@uchicago.edu}
     
\maketitle

\begin{abstract}
The age of the Universe at a given redshift is a fundamental relationship in cosmology.  For many years, the uncertainties in it were dauntingly large, close to a factor of 2.  In this age of precision cosmology, they are now at the percent level and dominated by the uncertainty in the Hubble constant.  The uncertainties due to the parameters that describe the current cosmological model, $\Lambda$CDM, are much less important.  In decreasing order they are:  uncertainty due to the matter density $\Omega_M$ around 0.9\% for $z> 3$; uncertainty due to the dark energy equation-of-state parameter $w$, less than 0.3\% for $z>3$;   and  uncertainty due to the curvature parameter $\Omega_k$, at most 0.07\%.


\end{abstract}
\vskip 20pt

With only the assumption of the Robertson-Walker metric it is simple to show that the age of the Universe at a given redshift is given by (see e.g., Ref.~\cite{KolbTurner})
\begin{eqnarray}
t(z) & = & \int_z^\infty {dz \over (1+z) H(z)} \nonumber \\
 & = & H_0^{-1}\int_z^\infty {dz \over (1+z) \left[ (1-\Omega_M+\Omega_k) (1+z)^{3(1+w)} -\Omega_k (1+z)^2 +\Omega_M (1+z)^3 \right]^{1/2} } \nonumber \\
& = & H_0^{-1} \int_{(1+z)}^\infty  {dx\over x \left[ (1-\Omega_M+\Omega_k) x^{3(1+w)} -\Omega_k x^2 +\Omega_M x^3 \right]^{1/2} }
\end{eqnarray}
where $x\equiv 1+z$, $\Omega_M$ is the fraction of critical density in matter, $w$ is the dark energy equation-of-state parameter, and $\Omega_k$ is the curvature parameter.  Here it is assumed that the Universe comprises matter and dark energy and the small, about 0.01\% fraction of critical density in radiation has been neglected here (see footnote below).  

Twenty-five years ago, the uncertainties in the age-redshift were very large, both because of uncertainty in the Hubble constant and in the cosmological model.  Until 2001, when the HST Key Project pinned down $H_0$ with a reliable error budget \cite{HSTKeyProject}, one could argue for any value between 50\,km/sec/Mpc and 100\,km/sec/Mpc (or even beyond \cite{Bartlettetal}) and the use of ``little $h$'' was universal, $h \equiv H_0/$100/km/sec/Mpc.  Likewise, values for the density parameter $\Omega_0$ ranged from $0.1$ to $1.0$ and even above.  Together, this meant that $t(z)$ was uncertain by around a factor of two, making quantifying the uncertainty in the age-redshift relationship a fool's errand.

In the current era of precision cosmology \cite{TurnerARNPS}, the situation couldn't be more different:  We have a standard model, $\Lambda$CDM, well supported by data, with a small set of cosmological parameters that are well determined.  Likewise, the Hubble constant is known to a precision of better than 5\% \cite{Riess,Freedman}. For $\Lambda$CDM, $\Omega_k = 0$ and $w=-1$, the age-redshift relationship is given by:
$$ t(z) =  {1\over 3}H_0^{-1}\Omega_\Lambda^{-1/2} \ln \left[ {[1+ x^3\Omega_M/\Omega_\Lambda ]^{1/2} +1 \over [1+ x^3\Omega_M/\Omega_\Lambda ]^{1/2} -1} \right] \nonumber \\ $$
where as before $x=1+z$.

The purpose of this note is to estimate the uncertainty in the age-redshift relationship for the $\Lambda$CDM model.  In particular, with the exception of $H_0$ (see below), I will be guided by the cosmological parameters and uncertainties from Planck \cite{PlanckLegacy}, supplemented by measurements of $w$ by Pantheon \cite{Pantheon} and the Dark Energy Survey \cite{DES5}:
\begin{itemize}
\item $\Omega_M = 0.3111\pm 0.0056$
\item $\Omega_\Lambda = 0.6889 \pm 0.0056$
\item $\Omega_k = 0.0007 \pm 0.0019$
\item $w=-1 \pm 0.05$
\end{itemize}
where I have inflated the quoted uncertainty in $w$ by a factor of 2.

Adopting these cosmological parameters, i.e., $\Omega_M = 0.3111$, $\Omega_k = 0$ and $w = -1$, the age-redshift relationship above is
\begin{eqnarray}
t(z) & = & 0.402H_0^{-1} \ln \left[ {(1+0.452 x^3)^{1/2} + 1 \over (1+0.452 x^3)^{1/2} - 1 } \right] \nonumber \\
& = & 5.616 (0.7/h)\,{\rm Gyr} \ln \left[ {(1+0.452 x^3)^{1/2} + 1 \over (1+0.452 x^3)^{1/2} - 1} \right]
\end{eqnarray}

The determination of the Hubble constant is once again of pressing interest and controversy, with apparent discrepancies between the value directly measured and that inferred from CMB anisotropy measurements and the assumption of $\Lambda$CDM \cite{Riess,Freedman}.  Since the Hubble constant factors out in determining the age-redshift relationship it is easy to separate its contribution to the uncertainty in $t(z)$ from the other cosmological parameters, remembering that there may be correlations with the other parameters.  Without making a judgment on the current Hubble tension, and for our purposes here, I think it is fair to say the current uncertainty is in the range of 1\% to 5\%.

Taking as the independent cosmological parameters, $H_0$, $\Omega_M$, $w$ and $\Omega_k$, it follows by differentiation that
\begin{eqnarray}
{\delta t \over t(z)} & = & -{\delta H_0 \over H_0} \\
{\delta t \over \delta \Omega_M} &= & -{1\over 2}  H_0^{-1} \int_{1+z}^\infty { (x^3 -1 ) dx \over x \left[ (1-\Omega_M+\Omega_k) x^{3(1+w)} -\Omega_k x^2 +\Omega_M x^3 \right]^{3/2} } \\
{\delta t \over \delta w} &= & -{3\over 2}  H_0^{-1} \int_{1+z}^\infty { \ln (x)(1-\Omega_M+\Omega_k) x^{3(1+w)}dx \over x \left[ (1-\Omega_M+\Omega_k) x^{3(1+w)} -\Omega_k x^2 +\Omega_M x^3 \right]^{3/2} } \\
{\delta t \over \delta \Omega_k} & =& {1\over 2} H_0^{-1} \int_{1+z}^\infty { (x^2 -1 ) dx \over x \left[ (1-\Omega_M+\Omega_k) x^{3(1+w)} -\Omega_k x^2 +\Omega_M x^3 \right]^{3/2} }
\end{eqnarray}

Bringing everything together, the uncertainty in the age-redshift relationship can be written as:
$$ {\delta t \over t(z)} =  \left\{ -{\delta H_0 / H_0}, \  \alpha_M \delta\Omega_M, \  \alpha_w \delta_w, \  \alpha_k \delta\Omega_k \right\}, $$
where
\begin{eqnarray}
\alpha_M & = & -{1\over 2\alpha_0} \int_{1+z}^\infty {(x^3 -1 ) dx \over x \left[ (1-\Omega_M+\Omega_k) x^{3(1+w)} -\Omega_k x^2 +\Omega_M x^3 \right]^{3/2} } \\
\alpha_w & = & -{3\over 2 \alpha_0} \int_{1+z}^\infty {\ln (x)(1-\Omega_M+\Omega_k) x^{3(1+w)} dx \over x \left[ (1-\Omega_M+\Omega_k) x^{3(1+w)} -\Omega_k x^2 +\Omega_M x^3 \right]^{3/2} } \\
\alpha_k & = & {1\over 2 \alpha_0} \int_{1+z}^\infty {(x^2 -1 ) dx \over x \left[ (1-\Omega_M+\Omega_k) x^{3(1+w)} -\Omega_k x^2 +\Omega_M x^3 \right]^{3/2} } \\
\alpha_0 & = & \int_{1+z}^\infty { dx \over x \left[ (1-\Omega_M+\Omega_k) x^{3(1+w)} -\Omega_k x^2 +\Omega_M x^3 \right]^{1/2} } 
\end{eqnarray}
I have not added the uncertainties in quadrature, which would require a choice for the $H_0$ uncertainty and taking proper account of correlated error in the cosmological parameters.  Rather, I have shown the individual contributions separately to illustrate their relative importance and to allow the reader to re-scale if they feel it is necessary.

Fig.~1 shows the relative uncertainties in the age-redshift relationship as a function of redshift.  From the least important to the most important they are: $\Omega_k$, at most 0.07\%; $w$, at most 0.2\% for $z>3$; $\Omega_M$, around 0.9\% for $z>3$; and $H_0$, between 1\% and 5\% (which follows directly from its assumed uncertainty).\footnote{Had one included the radiation component, $\Omega_R\simeq 0.00008$ whose major uncertainty owes to that in $H_0$, its contribution to the uncertainty in the age-redshift relationship would be at most 0.01\%, and of course, correlated with $H_0$.}  

The redshift dependence of the uncertainty upon $\Omega_M$ and $w$ is easy to understand.  It is most dependent upon $w$ at $z<1$ where dark energy dominates and the age is close to that of a dark energy dominated universe.  The uncertainty due to $w$ drops rapidly with redshift because for $z>1$ the Universe is increasingly matter-dominated and the age approaches that of a matter only universe.  In particular, $\delta t (z)/t(z) \rightarrow -{1\over 2}\delta\Omega_M/\Omega_M  = 0.93\%$ for $z\gg 1$..

\begin{figure}[h]
\center\includegraphics[width = 0.7\textwidth]{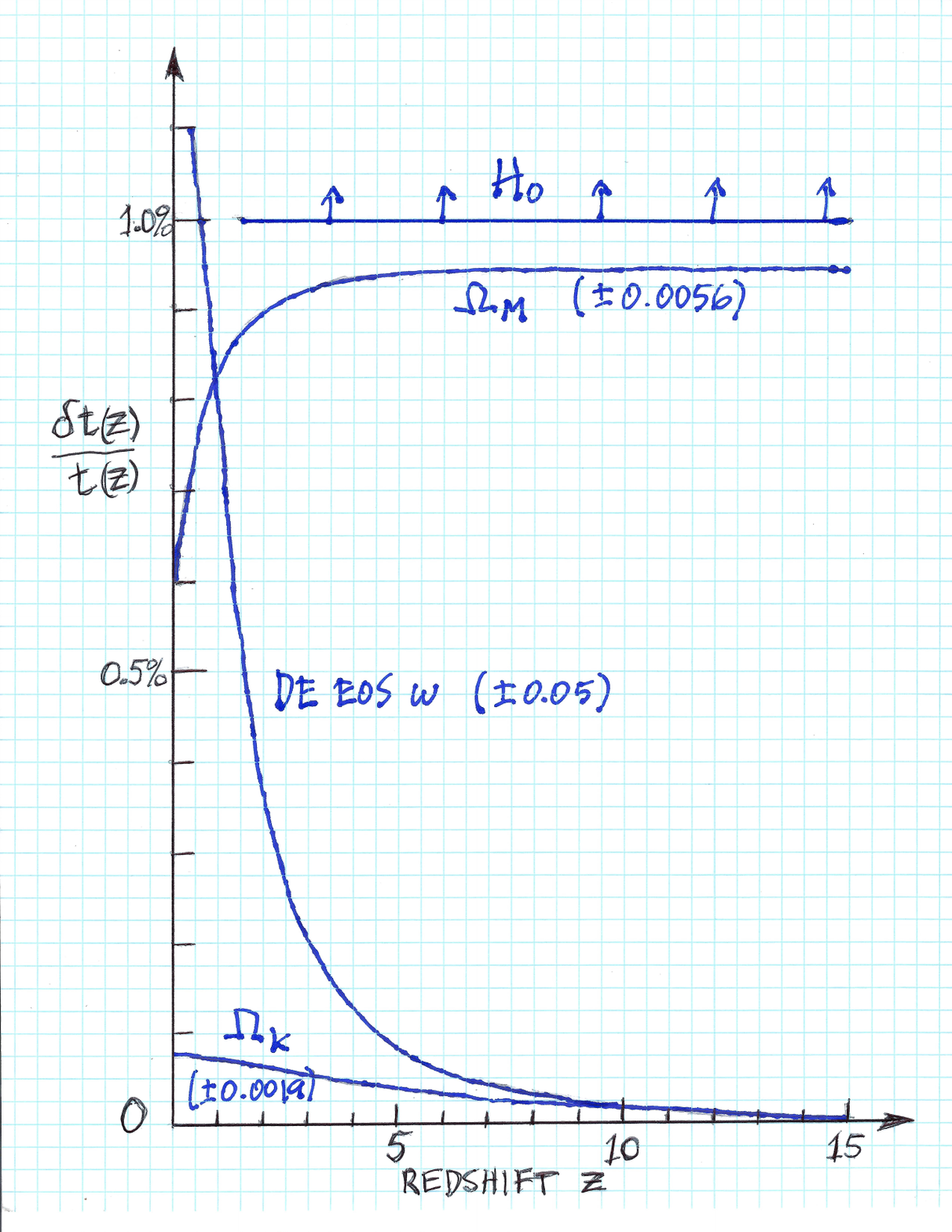}
\caption{Contributions to the uncertainty in the age of the Universe at redshift $z$ from the cosmological parameters $H_0$, $w$, $\Omega_M$ and $\Omega_k$.}
\end{figure}

For completeness, now consider the look-back time:
\begin{eqnarray}
t_{LB}(z) & \equiv & \int_0^z {dz \over (1+z) H(z)} \nonumber \\
 & = & H_0^{-1}\int_0^z {dz \over (1+z) \left[ (1-\Omega_M+\Omega_k) (1+z)^{3(1+w)} -\Omega_k (1+z)^2 +\Omega_M (1+z)^3 \right]^{1/2} } \nonumber \\
& = & H_0^{-1} \int_1^{(1+z)} {dx\over x \left[ (1-\Omega_M+\Omega_k) x^{3(1+w)} -\Omega_k x^2 +\Omega_M x^3 \right]^{1/2} }
\end{eqnarray}

The analogous formulae for the uncertainty in it are:

\begin{eqnarray}
{\delta t_{LB} \over t_{LB}(z)} & = & -{\delta H_0 \over H_0} \\
{\delta t_{LB}\over \delta \Omega_M} &= & -{1\over 2}  H_0^{-1} \int_1^{1+z} { (x^3 -1 ) dx \over x \left[ (1-\Omega_M+\Omega_k) x^{3(1+w)} -\Omega_k x^2 +\Omega_M x^3 \right]^{3/2} } \\
{\delta t_{LB} \over \delta w} &= & -{3\over 2}  H_0^{-1} \int_1^{1+z} { \ln (x)(1-\Omega_M+\Omega_k) x^{3(1+w)}dx \over x \left[ (1-\Omega_M+\Omega_k) x^{3(1+w)} -\Omega_k x^2 +\Omega_M x^3 \right]^{3/2} } \\
{\delta t_{LB} \over \delta \Omega_k} & =& {1\over 2} H_0^{-1} \int_1^{1+z} { (x^2 -1 ) dx \over x \left[ (1-\Omega_M+\Omega_k) x^{3(1+w)} -\Omega_k x^2 +\Omega_M x^3 \right]^{3/2} }
\end{eqnarray}
$$ {\delta t_{LB} \over t_{LB}(z)} =  \left\{ -{\delta H_0 / H_0}, \  \beta_M \delta\Omega_M, \  \beta_w \delta_w, \  \beta_k \delta\Omega_k \right\}, $$
where
\begin{eqnarray}
\beta_M & = & -{1\over 2\beta_0} \int_1^{1+z}  {(x^3 -1 ) dx \over x \left[ (1-\Omega_M+\Omega_k) x^{3(1+w)} -\Omega_k x^2 +\Omega_M x^3 \right]^{3/2} } \\
\beta_w & = & -{3\over 2 \beta_0} \int_1^{1+z} {\ln (x)(1-\Omega_M+\Omega_k) x^{3(1+w)} dx \over x \left[ (1-\Omega_M+\Omega_k) x^{3(1+w)} -\Omega_k x^2 +\Omega_M x^3 \right]^{3/2} } \\
\beta_k & = & {1\over 2 \beta_0} \int_1^{1+z} {(x^2 -1 ) dx \over x \left[ (1-\Omega_M+\Omega_k) x^{3(1+w)} -\Omega_k x^2 +\Omega_M x^3 \right]^{3/2} } \\
\beta_0 & = & \int_1^{1+z} { dx \over x \left[ (1-\Omega_M+\Omega_k) x^{3(1+w)} -\Omega_k x^2 +\Omega_M x^3 \right]^{1/2} } 
\end{eqnarray}
Fig.~2 shows the relative uncertainties in the look-back time as a function of redshift.  From the least important to the most important they are: $\Omega_k$, at most 0.07\%; $\Omega_M$, at most 0.5\%; $w$, at most 0.9\%; and $H_0$, between 1\% and 5\%.  Here too the dependence of the uncertainties is easy to understand.  For small redshifts, $z < 1$, the uncertainty due to knowledge of $\Omega_M$ is less important because look back only encompasses the dark energy dominated era.  The uncertainty due to $\Omega_M$ rises at higher redshift because  look back includes the matter-dominated era as well. 

\begin{figure}[h]
\center\includegraphics[width = 0.7\textwidth]{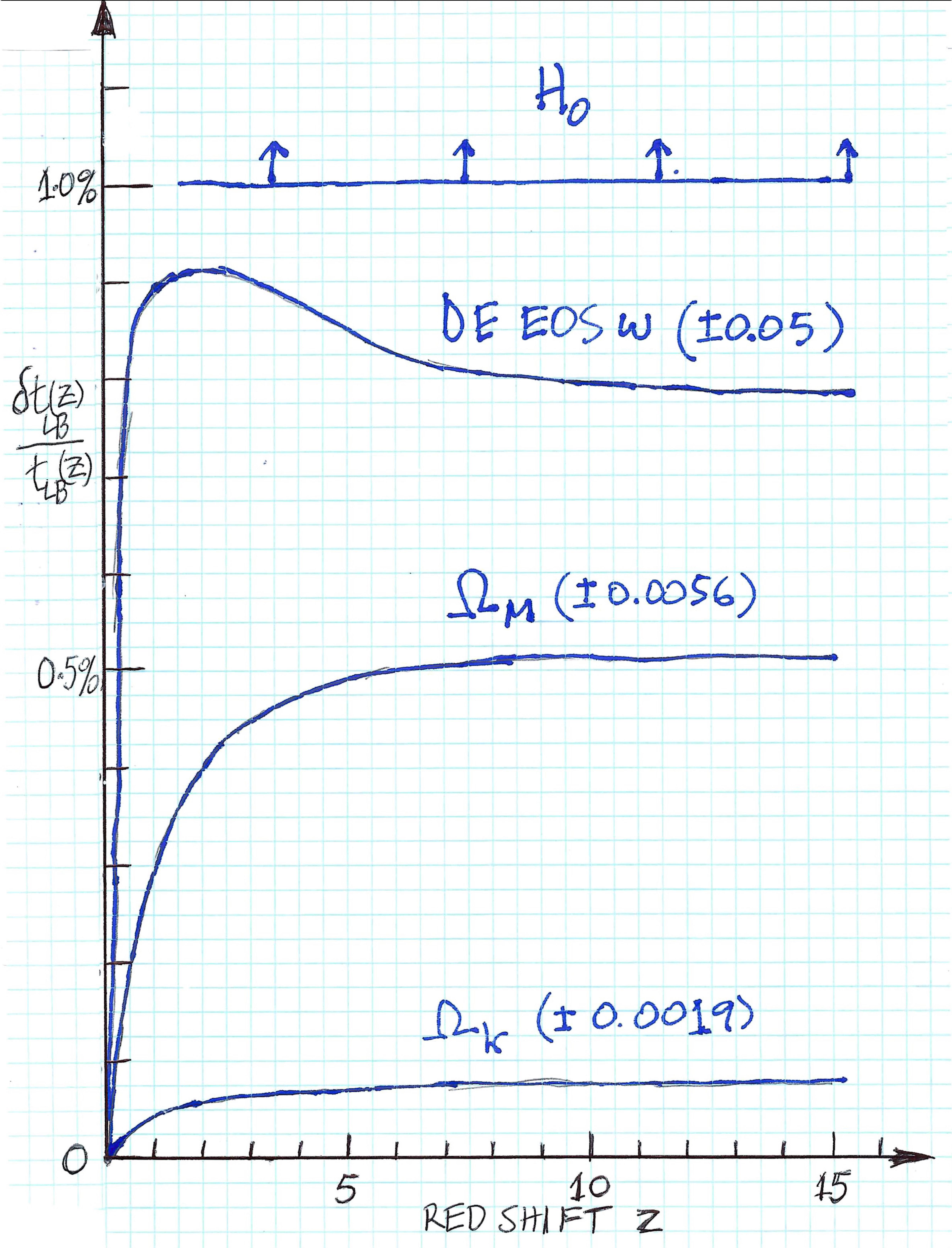}
\caption{Contributions to the uncertainty in the look back time at redshift $z$ from the cosmological parameters $H_0$, $w$, $\Omega_M$ and $\Omega_k$.}
\end{figure}

Two final comments.  First, other than a variation in the dark energy EOS parameter, I did not consider modifications to the $\Lambda$CDM paradigm.  The effect of early dark energy on the age-redshift relationship was discussed in detail in Ref.~\cite{Weisz}.  As I mentioned above, I did not combine all the parameter uncertainties,  because of the correlations between the values determined for the parameters.  But more importantly, the Hubble tension raises the issue of a possible missing element in the current cosmological model or a systematic error in $H_0$, which could impact $t(z)$ significantly.

\hskip 20 pt

This note was inspired by a question asked of my UCLA colleague Professor Alice Shapley at the International Space Science Institute's {\it Symposium on Cosmic Beginnings:  Unveiling the first billion years with JWST} at the University of Bern on 14 March 2024.  The question was, what is the uncertainty in the age-redshift relationship.  She responded that it was small but she didn't know precisely how small.  I thought it would make a nice problem for my UCLA Cosmology class' final problem set.  Unfortunately, I had just given out that problem set and so I did the problem myself to share with the class.



\end{document}